\documentstyle[psfig,letters]{mn}

\newif\ifAMStwofonts

\def\dfe{{$\frac{df}{d\epsilon}$}\ }
\def\fbar{{$\bar{f}_r$}\ }

\def\mcg{{MCG$-$6-30-15}\ }
\def\ttm{{T $<$ T$_m$}}
\def\xh{{X$_H$}\ }
\def\xli{{X$_{Li}$}\ }
\def\efe25{{$(e~+~Fe~XXV) \rightarrow Fe~XXIV$}\ }
\def\ka{{K$\alpha$}\ }
\def\etal{{\it et\thinspace al.}\ }

\def\lam{{$\lambda$}\ }

\def\eion{{(e~+~ion)}\ }
\newcommand{\be}{\begin{equation}}
\newcommand{\ee}{\end{equation}}

\ifoldfss
  \ifCUPmtlplainloaded \else
    \NewTextAlphabet{textbfit} {cmbxti10} {}
    \NewTextAlphabet{textbfss} {cmssbx10} {}
    \NewMathAlphabet{mathbfit} {cmbxti10} {} 
    \NewMathAlphabet{mathbfss} {cmssbx10} {} 
  \fi
  \ifAMStwofonts
    \ifCUPmtlplainloaded \else
      \NewSymbolFont{upmath} {eurm10}
      \NewSymbolFont{AMSa} {msam10}
      \NewMathSymbol{\upi}     {0}{upmath}{19}
      \NewMathSymbol{\umu}     {0}{upmath}{16}
      \NewMathSymbol{\upartial}{0}{upmath}{40}
      \NewMathSymbol{\leqslant}{3}{AMSa}{36}
      \NewMathSymbol{\geqslant}{3}{AMSa}{3E}

      \let\leq=\leqslant 
      \let\geq=\geqslant 
    \fi
  \fi
\fi 

\ifnfssone
  \newmathalphabet{\mathit}
  \addtoversion{normal}{\mathit}{cmr}{m}{it}
  \addtoversion{bold}{\mathit}{cmr}{bx}{it}
  \newmathalphabet{\mathbfit} 
  \addtoversion{normal}{\mathbfit}{cmr}{bx}{it}
  \addtoversion{bold}{\mathbfit}{cmr}{bx}{it}
  \newmathalphabet{\mathbfss} 
  \addtoversion{normal}{\mathbfss}{cmss}{bx}{n}
  \addtoversion{bold}{\mathbfss}{cmss}{bx}{n}
  \ifAMStwofonts
    \ifCUPmtlplainloaded \else
      \UseAMStwoboldmath
      \makeatletter
      \new@mathgroup\upmath@group
      \define@mathgroup\mv@normal\upmath@group{eur}{m}{n}
      \define@mathgroup\mv@bold\upmath@group{eur}{b}{n}
      \edef\UPM{\hexnumber\upmath@group}
      \new@mathgroup\amsa@group
      \define@mathgroup\mv@normal\amsa@group{msa}{m}{n}
      \define@mathgroup\mv@bold\amsa@group{msa}{m}{n}
      \edef\AMSa{\hexnumber\amsa@group}
      \makeatother
      \mathchardef\upi="0\UPM19
      \mathchardef\umu="0\UPM16
      \mathchardef\upartial="0\UPM40
      \mathchardef\leqslant="3\AMSa36
      \mathchardef\geqslant="3\AMSa3E

      \let\leq=\leqslant 
      \let\geq=\geqslant 
    \fi
  \fi
\fi 

\ifnfsstwo
  \DeclareMathAlphabet{\mathbfit}{OT1}{cmr}{bx}{it}
  \SetMathAlphabet\mathbfit{bold}{OT1}{cmr}{bx}{it}
  \DeclareMathAlphabet{\mathbfss}{OT1}{cmss}{bx}{n}
  \SetMathAlphabet\mathbfss{bold}{OT1}{cmss}{bx}{n}
  \ifAMStwofonts
    \ifCUPmtlplainloaded \else
      \DeclareSymbolFont{UPM}{U}{eur}{m}{n}
      \SetSymbolFont{UPM}{bold}{U}{eur}{b}{n}
      \DeclareSymbolFont{AMSa}{U}{msa}{m}{n}
      \DeclareMathSymbol{\upi}{0}{UPM}{"19}
      \DeclareMathSymbol{\umu}{0}{UPM}{"16}
      \DeclareMathSymbol{\upartial}{0}{UPM}{"40}
      \DeclareMathSymbol{\leqslant}{3}{AMSa}{"36}
      \DeclareMathSymbol{\geqslant}{3}{AMSa}{"3E}

      \let\leq=\leqslant 
      \let\geq=\geqslant 
    \fi
  \fi
\fi 

\ifCUPmtlplainloaded \else
  \ifAMStwofonts \else 
    \def\upi{\pi}
    \def\umu{\mu}
    \def\upartial{\partial}
  \fi
\fi

\title{The K$\alpha$ complex of He-like iron with dielectronic
satellites}
\author[Justin Oelgoetz and Anil K. Pradhan]
       {Justin Oelgoetz$^1$ and Anil K. Pradhan$^2$\\
       $^1$ Department of Chemistry, $^2$ Department of Astronomy,
 The Ohio State University, Columbus, OH 43210, USA}
\date{Accepted  xxxxxx 
      Received xxxxxx;
      in original form xxxxxx}

\pagerange{\pageref{firstpage}--\pageref{lastpage}}
\pubyear{1994}

\def\LaTeX{L\kern-.36em\raise.3ex\hbox{a}\kern-.15em
    T\kern-.1667em\lower.7ex\hbox{E}\kern-.125emX}

\begin{document}

\maketitle

\label{firstpage}

\begin{abstract}

 It is shown that the dielectronic satellites (DES) dominate X-ray
spectral formation in the 6.7 keV \ka complex of Fe~XXV at temperatures below
that of maximum abundance in collisional ionization equilibrium T$_m$.
Owing to their extreme temperature sensitivity the DES
are excellent spectral diagnostics for T $<$ T$_m$ in
photoionized, colllisional, or hybrid plasmas; whereas the 
forbidden, intercombination, and resonance lines of Fe~XXV are not. 
A diagnostic line ratio GD(T) is defined 
including the DES and the lines, with parameters from new relativistic
atomic calculations. The DES {\it absorption}
resonance strengths may be obtained from differential oscillator 
strengths to possibly yield the Fe~XXIV/Fe~XXV column densities. 
The DES contribution to highly ionized Fe should be of
interest for models of redward broadening of \ka features,
ionized accretion discs, accretion flows, 
and \ka temporal-temperature variability in AGN.

\end{abstract}


\section{INTRODUCTION}
 The \ka features are complexes of lines from K-shell transitions.
Fe \ka lines are prominent in the X-ray spectra of AGN, with
associated photon energies that range from the so called `neutral iron'
6.4 keV fluorescent feature from Fe~I-Fe~XVI, up to 6.7 and 6.9 keV
from highly ionized He- and H-like Fe~XXV and Fe~XXVI.
The observed \ka
features from a number of AGN and quasars appear to peak at these
energies (e.g. Nandra \etal 1997).
The highly ionized
Fe~XXVI/Fe~XXV X-ray emission may serve as possible diagnostics of 
accretion flows and the structure of the source plasma 
(Naryayan and Raymond 1999). The nature of the accretion flow on to the
black hole or neutron star may 
determine whether the plasma is photoionization or collisionally
dominated, with distinct X-ray signatures. 
The X-ray spectra of a number of AGN exhibit a long redward tail
down to almost 5 keV (e.g. Tanaka \etal 1995, Nandra \etal 1997),
attributed to relativistic broadening of the emitting
radiation from the close proximity to the black hole. 
Accretion disc models assumed an essentially `neutral' plasma, with
a peak at 6.4 keV. The new {\it Chandra} observations show 
a narrow \ka feature of ionized Fe from 
the well known AGN NGC 5548 (Yaqoob \etal 2001).
Recently models have been proposed
with ionized Compton reflection from the inner region of the accretion disc,
and including separate diskline components (Ballantyne
and Fabian 2001). The 6.7 keV complex of Fe~XXV may constitute part of
these highly ionized Fe components. 
X-ray variability of Fe \ka emission has also been observed, 
for example, on rapid timescales in NGC 7314 
(Yaqoob \etal 1996), but constant over long time intervals from the AGN
\mcg6 with some evidence of short time scale variations
indicating possible flaring activity (Lee \etal 2000). 
As described in this work, the \ka complex
is sensitive to both the temperature and the ionization state, but the
spectral diagnostics needs to include the dielectronic satellites.
It is of interest to 
study precisely the basic physics of spectral formation of highly
ionized Fe that might discriminate between the atomic
and the astrophysical processes.

As first noted by Gabriel and Jordan (1969), and followed by other
studies (e.g. Mewe and Schrijver 1978a,b, Pradhan and Shull 1981,
Bely-Dubau \etal 1982), He-like ions provide 
powerful X-ray diagnostics. There is a
set of forbidden (f), intercombination (i), and resonance (r) lines
dependent on plasma parameters such as temperature, density, and
ionization state. The spectroscopic and laboratory designations for
these lines are: w,x,y,z, corresponding to the 4 transitions to the
ground level $1s^2 \ (^1S_0)
\longleftarrow 1s2p (^1P^o_1), 1s2p (^3P^o_2), 1s2p (^3P^o_1), 1s2s
(^3S_1)$ respectively. For temperatures \ttm, in recombination dominated
non-coronal plasmas it has been shown that \eion recombination
preferentially enhances the triplet (x,y,z) lines and hence the ratio G =
(x+y+z)/w, which may be thereby employed as diagnositcs of both
the temperature and the ionization state (Pradhan 1982, 1985). 
Since recombinations depend on
the ionization fraction of H-like to He-like ratio \xh = Fe~XXVI/Fe~XXV,
recombination dominated photoionized plasmas may be accordingly modeled
(Liedahl 2000, Porquet and Dubau 2000, Bautista and Kallman 2000). 

 The dielectronic satellites (DES) are generally formed due to {\it emission}
from autoionizing levels -- an excited ion level with a quasi-bound
electron. The \eion system may autoionize in a radiationless
transition, or stabilise (dieletronic recombination) 
via radiative decay with a photon
of wavelength redward of the `resonance' line (w) at 6.696 keV.
 Recently X-ray photoabsorption in KLL resonances of Li-like O~VI was
predicted (Pradhan 2000), and detected in
the {\it Chandra} observations of \mcg (Lee \etal 2001). 
Together with other observed photoabsorption and emission lines redward
of the resonance lines of H- and He-like O, and Fe L-shell spectra,
this helps in the spectral discrimination between a `dusty warm
absorber' model and alternative models of
relativistically broadened \ka lines
(Branduardi-Raymont \etal 2001). The
O~VI and O~VII column densities were also determined from the same 
X-ray observations (Lee
\etal 2001) using the computed O~VI(KLL) resonance oscillator strengths (Pradhan
2000, Nahar \etal 2001a). The DES photoabsorption calculations have now 
been extended to the KLL resonances of (e~+~C~V) and (e~+~Fe~XXV)
(Nahar \etal 2001a). In this {\it Letter} we examine the role
of DES in the \ka complex of Fe~XXV, both in emission and absorption,
and point out potentially useful diagnostics.

\section{THEORY AND COMPUTATIONS}

 A brief description of the characterstics of the principal line ratios 
and the DES based on previous laboratory, astrophysical, and theoretical
studies is given below.

\subsection{The principal lines w,x,y,z}

 All He-like ions provide diagnostic line ratios involving the principal lines from the bound levels of the ground and n = 2
configurations, R = z/(x+y) and G = (x+y+z)/w, that usually depend 
on density and temperature respectively. Furthermore, one can establish 
numerical limits on the ratio G that indicate the ionization state of the 
plasma:
 G $\approx 1  \rightarrow $  collisional (coronal) ionization equilibrium; 
G $> 1  \rightarrow $ recombination dominated (photoionization or
hybrid); G $< 1  \rightarrow $ ionization dominated. 
 Among the first examples of
observed non-ionization equilibrium plasmas
were: the recombination dominated laboratory tokamak 
spectra by K\"{a}llne \etal (1984), the {\it Einstein} X-ray spectra of 
ionization dominated plasmas in SNR's Puppis A 
(Winkler \etal 1981) and the Cygnus Loop (Vedder \etal 1986), and in
the laboratory tokamak spectra under rapid time-dependent
ionization (Lee \etal 1985,1986).

\subsection{The dielectronic satellites KLL and KLn ($n > 2$)}

Unlike elements lighter than iron, the
DES in Fe are comparably strong relative to the principal lines, even
for T $\approx$ T$_m$ in collisional equilibrium, 
such as in tokamak plasmas
(Bitter \etal 1981), in Electron Beam Ion Traps  
(EBIT, Beirsdorfer \etal 1992), 
and in the solar corona and flares (e.g. Kato \etal
1998). The autoionizing levels responsible for DES are excited
by electrons at precisely the energies corresponding to those levels. 
The DES are therefore more sensitive to the electron distribution (Maxwellian
or non-Maxwellian) than the principal lines which
are excited by all electrons at energies above the excitation threshold.
Moreover, the DES are spread over a much wider range in energy. 

 Following Gabriel (1972), several
researchers have computed the DES intensities (e.g. Bhalla \etal 1975, 
Bely-Dubau \etal 1979, 1982, Karim and Bhalla 1992, Vainshtein and
Safronova 1978).
To our knowledge, the most recent calculations of DES intensities
is by Pradhan and Zhang (1997), from highly resolved fine structure
photoionization cross sections
of Fe~XXIV including radiation damping of autoionizing resonances (Nahar
\etal 2001a).
These constitute a part of new relativistic atomic calculations for
photoionization, recombination, and
transition probabilities of Fe~XXV
(Zhang \etal 1999, Nahar and Pradhan 1999, Nahar \etal 2001b)
using the Breit-Pauli R-matrix (BPRM) method
(Hummer \etal 1993, Berrington \etal 1995).
 Photoionization and recombination are
treated self-consistently using an identical eigenfunction expansion for the
system (e~+~Fe~XXV) $\rightleftharpoons$ Fe~XXIV for the two inverse 
processes, based on a unified method for \eion
recombination that subsumes both the radiative and the dielectronic
recombination (RR and DR) processes in an ab initio manner (e.g.
Zhang \etal 1999).
The computed autoioinization and radiative rates for
Fe~XXV are given in Table 1 of Pradhan and Zhang (1997) for all DES of the 
$n$ = 2 KLL 
(1s2s2p) complex, labelled according to the standard notation `a-v' and
their spectroscopic designations (Gabriel 1972). 
The higher-$n$ DES, $3 \leq n \leq 10$
(KLM, KLN,.....,KLT) are also similarly calculated (the $n > 2$
contribution is negligible). The BPRM DES(KLL) strengths agree to
within 10 - 20\% of the EBIT experimental measurements, 
and with other works; the DES(KLn, $n > 2$) are
in similar agreement with Bely-Dubau \etal (1979).
Recombination-cascade coefficients for (w,x,y,z) are from Mewe and 
Schrijver (1978), and collisional rates from Pradhan (1985);
collisional rates for ISE DES are from Kato \etal (1995).

\section{RESULTS AND DISCUSSION}

 While details of the calculations will be reported in a subsequent paper,
we describe briefly the main theoretical
results relevant to spectral analysis of observations with
blended principal and DES features.
 
\subsection{Line ratios and dielectronic satellites}

 Fig. 1 shows the \ka complex of Fe~XXV including the 
(w,x,y,z) lines and the DES(KLL) at a range of temperatures from Log (T)
= 6.8 - 7.8 (the line profiles are thermally broadened). The various 
panels in Fig. 1 reveal the large temperature
variations and the dominance of the DES intensities relative to the 
(w,x,y,z) up to T $\approx 3 \times 10^7$ K ($<$ T$_m$).
He-like  ions span the largest
temperature range of any ionization state; T$_m$ (Fe~XXV) $\approx
10^{3.5 - 5.0}$ in coronal equilibrium (Arnaud and Raymond 1992).
The DES may be further sub-divided according to inner-shell excitation (ISE)
from Li-like Fe~XXIV (dotted, Fig. 1), and \eion dielectronic
recombination (DR) of (e~+~Fe~XXV) (dash-dot, Fig. 1).
  For T $\geq$ T$_m$ the w-line
is predominant since the large fraction of electrons in the Maxwellian
tail contribute to its intensity relative to all other lines.

\begin{figure}
\centering
\psfig{figure=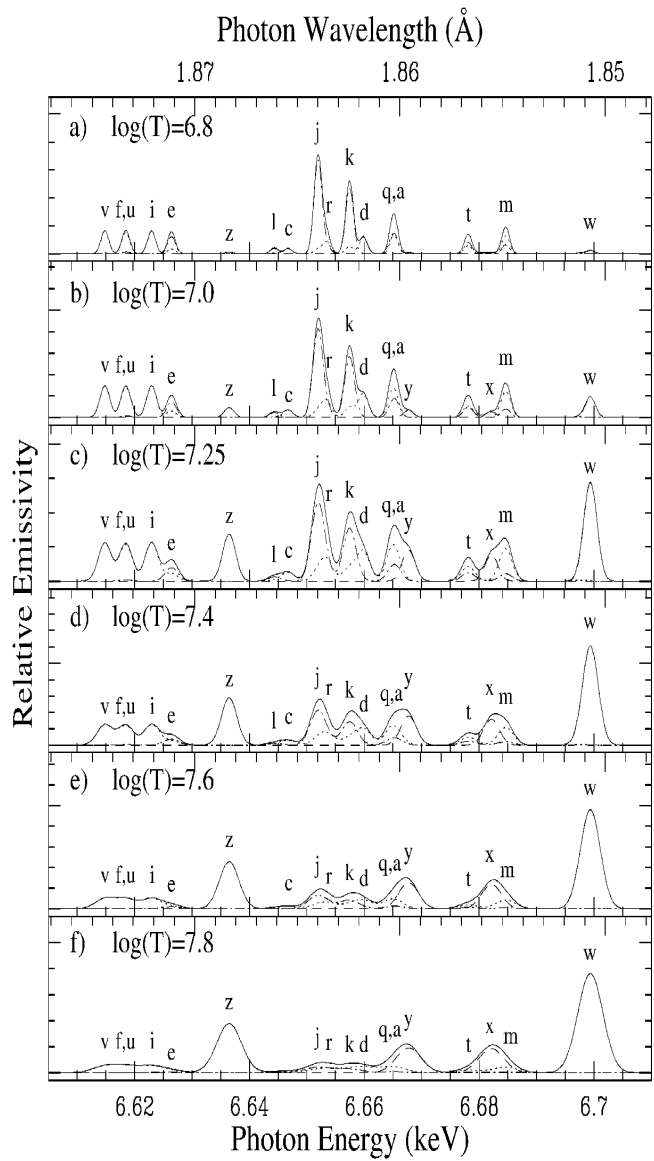,height=0.65\textheight,width=0.5\textwidth}
\caption{Fe~XXV(\ka) spectra vs. Log (T) = 6.8 - 7.8;
solid line - sum of principal lines w,x,y,z, and the DES(KLL);
dashes - (w,x,y,z); dotted - ISE satellites; dot-dash - DR satellites.
The features are thermally broadened.}
\end{figure}

 It is readily seen from Fig. 1 that the G ratio is no longer a
useful one for He-like iron. This is not only due to the relative weakness 
of the principal lines, but also blending with the DES
with different temperature dependences. Furthermore,
the contribution of $n \geq 3$ DES to the w-line needs to be considered.
We obtain the  higher-$n$ satellite intensities KLn ($ 2 < n \leq 10$, 
i.e. KLM, KLN, etc.), from the BPRM calculations as in Table 2 of 
Zhang \etal (1999). The DES(KLn, $n > 2$) can not be
resolved from the w-line by current X-ray observatories.
We assume that the DES(KLL) intensity may be measured, albeit
blended with the (x,y,z) lines, relative to the w-line
blended with the higher-$n$ satellites.
Therefore we re-define the G ratio, including DES, as

\be
    GD \equiv (x+y+z+KLL) \ / \ (w+ \sum_{n>2}KLn) 
\ee

 Fig. 2a shows GD(T) compared with G(T) in coronal equilibrium. 
Whereas the latter shows
practically no variation, the GD(T) increases by up to a factor of 7
compared to G(T) at the lower end of the temperature range $\approx 6
\times 10^6$ K, and compared to T $\geq$ T$_m$. Therefore in
photoionized plasmas, or in otherwise low-T recombination dominated
plasmas, the ratio GD(T) should be a good diagnostics of temperature
with a limiting value GD(T$<<$T$_m$) $\longrightarrow$ KLL/$\sum_{n>2}$ KLn.

\begin{figure}
\centering
\psfig{figure=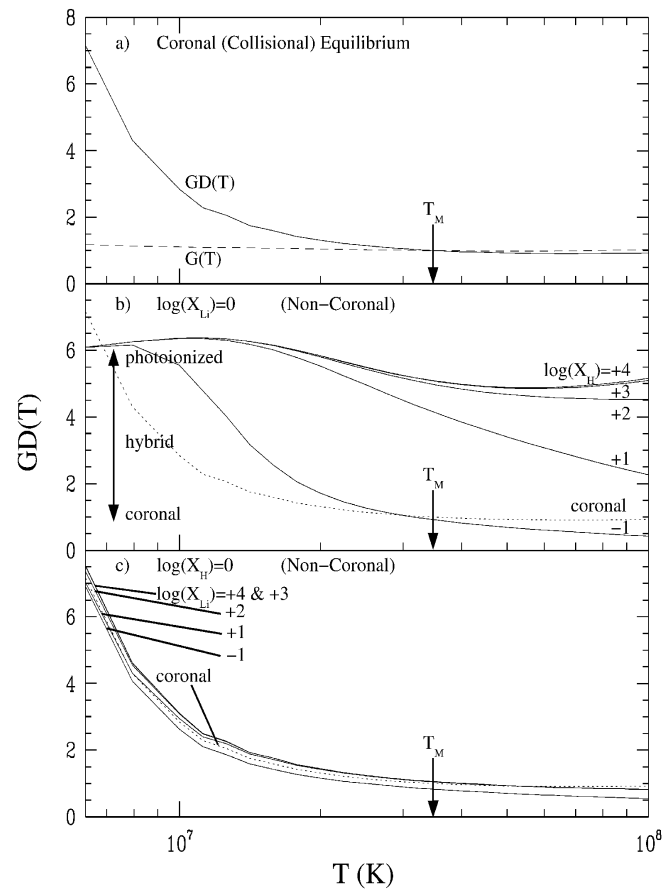,width=0.5\textwidth,height=0.5\textheight}
\caption{The ratio GD(T) in Eq. 1: a) in coronal equilibrium, compared
to G(T); b) non-coronal GD(T) variation in recombination dominated
plasmas with \xh = N(Fe~XXVI)/N(Fe~XXV); c) GD(T) variation with \xli
= N(Fe~XXIV)/N(Fe~XXV).}
\end{figure}

 The DES blending and dominance implies however that GD(T) will not
be as good a diagnostic of ionization state for He-like Fe as for lighter
elements. The DES are largely independent of ionization state; the
DR satellites depend only on Fe~XXV, and the
 relative intensities of (w,x,y,z) are smaller. The non-coronal
cases are illustrated in Figs. 2b and 2c.  Fig. 2b shows the
variation of GD(T) with H-like to He-like ionization fraction \xh in
completely recombination dominated plasmas (\xli $\rightarrow 0$).
Fig. 2c shows the
variation with respect to the Li-like to He-like ionization fraction \xli
in plasmas ionizing through  Fe~XXIV
(\xh $\rightarrow 0$). Unlike the recombination dominated case in Fig. 2b,
the temperature dependences in Figs. 2a (coronal equilibrium) and 2c are 
similar, since in both cases the DES
dominate at lower T, and collisional excitation of (w,x,y,z) at higher T.
The non-coronal recombination dominated case in Fig. 2b is similar to G(T) 
for all He-like ions, since recombinations
populate only the bound levels that lead preferentially to (x,y,z) emission,
 and not autoionizing levels that yield the DES. The curves in Fig. 2b
are therefore similar to that by Porquet and Dubau (2000) who considered
He-like ions of elements up to Si, but where the DES are relatively 
inconsequential.
 As in Fig. 2a, the Bautista and Kallman (2000) results for He-like Fe
including the DES also show a sharp rise in G(T) as T decreases; 
however they present
values only for G $\leq$ 2.5, and therefore it is 
difficult to discern wheteher there is quantitave
agreement with the present results.

 As with the G ratio, the density sensitive ratio R = z/(x+y) is not
likely to provide accurate diagnostics for Fe~XXV, since the DES blends 
with different types of ISE and DR satellites will introduce temperature and 
ionization-state dependence in R as well. Nonetheless, a more 
detailed analysis of all
components of the \ka complex (i.e. Fig. 1)  might enable calculations of the
R, including the DES, in cases with sufficient observational
resolution ($\Delta$ \lam $< 0.01 \AA$).

\subsection{The total \ka (Fe XXV) intensity}

The total intensity 
I(K$\alpha$) vs. T is plotted in  Fig. 3 (relative units),
showing the relative contributions of DES and principal lines:
I(KLL), I(KLn, n$>$ 2), and I(w+x+y+z). We note that: (i)
I (KLL) $\approx$ I(\ka) for T $\leq 10^7$ K, i.e. the DES(KLL) constitute
almost the entire \ka intensity, (ii) the DES (KLL + $\sum_{n>2}$KLn) 
continue to
dominate \ka up to 3 $\times 10^7$ K (and hence the temperature
dependence of GD(T) in Fig. 2a), and (iii) For T $>$ T$_m$ the DES are
weaker and the principal lines approach G(T).
The inset in Fig. 3 shows I(\ka) on a different scale, with the peak
values and the relative ionic abundances of Fe~XXIV, Fe~XXV, and Fe~XXVI
vs. T. The maxima of \ka
(Fe~XXV) components do not coincide with the Fe~XXV peak abundances 
N$_{Fe~XXV}$ (T$_m$); rather, the intensities also depend on adjacent
ionization states and the DES.
It bears emphasis that under non-equilibrium, time-dependent, 
 rapidly variable plasma conditions the line
ratios and the DES may vary widely in intensity.
Thus we expect a strong and sensitive temporal-temperature variation of the
entire \ka complex.

\begin{figure}
\centering
\psfig{figure=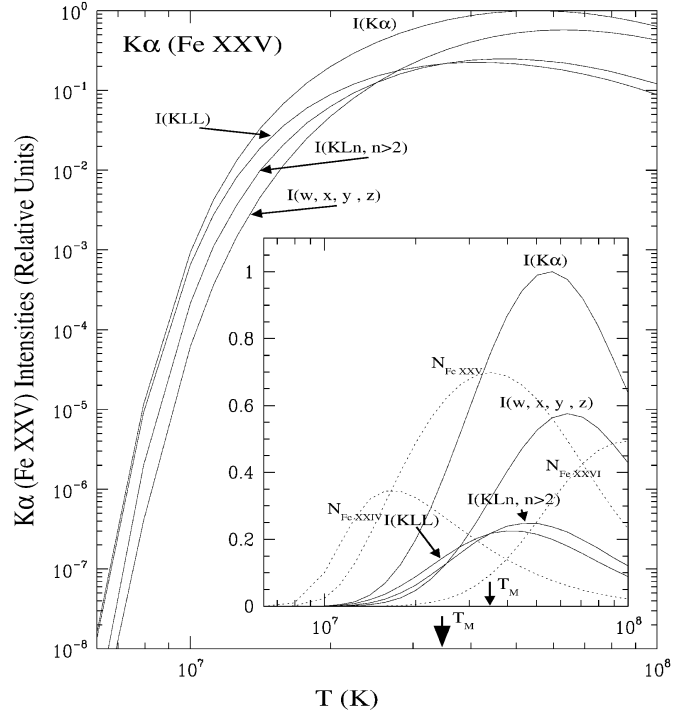,width=0.5\textwidth,height=0.4\textheight}
\caption{\ka (Fe~XXV) intensity, with I(KLL), I(KLn, n$>$2), and
I(w,x,y,z); the inset is on different scale, also showing relative
coronal equilibrium abundances. As shown, the DES dominate \ka
intensity for T $<$ T$_m$.}
\end{figure}

\begin{figure}
\centering
\psfig{figure=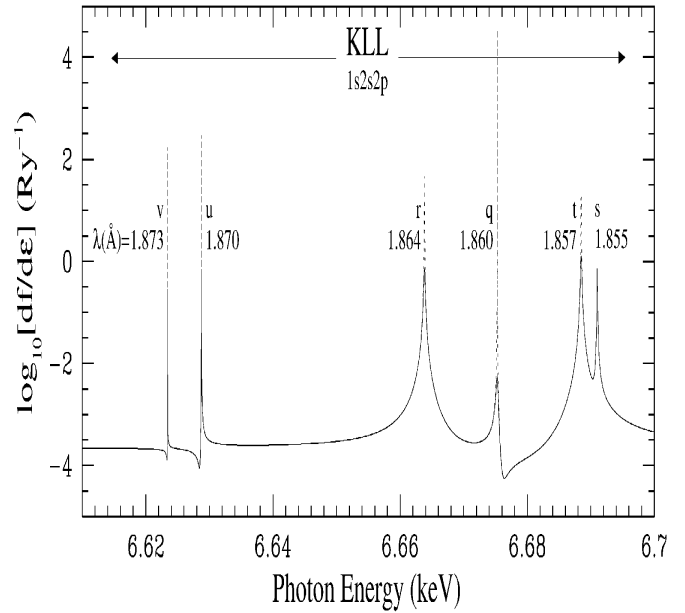,width=0.5\textwidth,height=0.35\textheight}
\caption{Fe~XXIV KLL differential oscillator strengths with resolved
X-ray spectral features (adapted from Nahar \etal 2001a). The soild
lines are radiatively damped values, and the dashed lines are
undamped values appropriate for optically thin plasmas.}
\end{figure}

\subsection{Absorption spectra of the DES(KLL)}

 Finally, in Fig. 4 we consider the {\it absorption} spectra in the KLL
satellites due to X-ray photoabsorption from
Fe~XXIV. As mentioned earlier, we have calculated differential
oscillator strengths, \dfe, for the KLL resonances in He-like C,O, and Fe
from highly resolved relativistic photoionization cross sections 
(Nahar \etal 2001a). Fig. 4 shows the detailed \dfe with the spectral
energy absorption across the ISE DES profiles.
Integration over the individual profiles yields the \fbar, the
resonance oscillator strengths that may be used in a completely analogous 
manner as the line $f$-values (Pradhan 2000). At 1.860 $\AA$ we may predict a
strong absorption feature due largely to the `q' ISE satellite (and to a
lesser extent from the `r' and `t'), with \fbar = 0.75 (0.044) (from
Table 1, Nahar \etal 2001a). The two \fbar values correspond to radiatively
undamped and damped values (dashed and solid lines); 
the former is appropriate for optically
thin plasmas with little re-emission along the line of sight. 
If the 1.86 $\AA$ feature is resolvable from the 
\lam(w) at 1.85 $\AA$ of Fe~XXV ($f$(w) = 0.703),
then, in principle, Fe~XXIV/Fe~XXV
column densities may be determined from measured equivalent
widths using curves-of-growth.
However, that may not be feasible due to y-line
emission also at 1.86 $\AA$, and other DES.
Nonetheless,
X-ray {\it absorption} in the DES (ISE) needs to be considered in
\ka spectral analysis.

\section{CONCLUSION}

 The main results are:

1. The highly ionized 6.7 keV Fe~XXV component of Fe \ka emission
is dominated by the DES. Those from H-like
Fe~XXVI should similarly contribute at $\sim$ 6.9 keV.
Resonance and satellite line emission from lower ionization states of Fe
should manifest itself down to 6.4 keV.

2. The DES appear redward of the w-line, and should be largely 
undiminished by resonance scattering
unlike the w-line; therefore the observed \ka features might appear
redshifted.

3. The principal line ratios G and R from the $n$ = 2 excited levels 
involving the
lines (w,x,y,z) are not reliable diagnostics for Fe~XXV due to DES 
strengths and blending. A new ratio GD(T) is defined including the DES.

4. Owing to the predominance of DES the \ka complex may exhibit
considerable temperature sensitivity, and hence possible temporal
variability in AGN.

5. K-shell DES {\it absorption} may be considered 
 using `resonance oscillator strengths' \fbar 
to determine column densities and ionic fractions.

\section*{Acknowledgments}
 We would like to thank Hong Lin Zhang for the data on high-$n$ DES,
and Sultana Nahar for the \dfe (KLL) data.
This work was partially supported by the U.S. National 
Science Foundation and NASA Astrophysical Theory Program. 
The computational work was carried out
at the Ohio Supercomputer Center in Columbus Ohio.

\def\amp{{ Adv. At. Molec. Phys.}\ }
\def\apj{{ApJ}\ }
\def\apjs{{ApJS}\ }
\def\aj{{AJ}\ }
\def\aa{{A\&A}\ }
\def\aasup{{A\&AS}\ }
\def\adndt{{ADNDT}\ }
\def\cpc{{Comput. Phys. Commun.}\ }
\def\jqsrt{{JQSRT}\ }
\def\jpb{{J. Phys. B}\ }
\def\pasp{{PASP}\ }
\def\mn{{MNRAS}\ }
\def\psc{{Phys. Scr.}\ }
\def\pra{{Phys. Rev. A}\ }
\def\prl{{Phys. Rev. Letts.}\ }

\clearpage





\label{lastpage}

\end{document}